\documentclass[amsmath,amssymb,pra]{revtex4}
\usepackage{graphicx}
\usepackage{amsfonts}
\usepackage{pst-text}
\usepackage{subfigure}

\newcommand{\bra}[1]{\langle{#1}|}
\newcommand{\ket}[1]{|{#1}\rangle}
\newcommand{\A}{\mathbb{A}}
\newcommand{\B}{\mathbb{B}}
\newcommand{\offset}{\textit{offset}}
\renewcommand{\mod}{\mathrm{mod}}
\newcommand{\Tr}{\mathrm{Tr}}

\newcommand{\IITISPAN}{The Institute of Theoretical and Applied Informatics of the Polish Academy of Sciences, Ba{\l}tycka 5, 44-100 Gliwice, Poland}
\newcommand{\USKATOWICE}{Institute of Physics, University of Silesia, Uniwersytecka 4, 40-007 Katowice, Poland}
\begin{document}

\title{Quantum Implementation of Parrondo's Paradox}

\author{Piotr Gawron\footnote{corresponding author: gawron@iitis.gliwice.pl}}
\affiliation{\IITISPAN}

\author{Jaros{\l}aw A.~Miszczak}
\affiliation{\USKATOWICE}
\affiliation{\IITISPAN}



\begin{abstract}
We propose a~quantum implementation of a~capital-dependent Parrondo's paradox that uses $O(\log_2(n))$ qubits, where $n$ is the number of Parrondo games. We present its implementation in the quantum computer language (QCL) and show simulation results. 
\end{abstract}

\maketitle

\section{Introduction}
Quantum game theory~\cite{PS03, PS04, flitabott-introduction} is a new field of science having its roots in both game theory and quantum information theory. For about a decade quantum computer scientists have been searching for new methods of quantum algorithm design. Thorough investigation of different quantum games may bring new insight into the development of quantum algorithms. 

It was shown that Grover's algorithm \cite{Gro96a} can be treated as an example of a quantum Parrondo's paradox \cite{LeJo02, LeJo02-2}. Operators used in Grover's algorithm can be treated as Parrondo games having separately zero expected values, however, if they are interwired, the expected value fluctuates. This effect is well known in Grover's algorithm. 

Implementation of a quantum Parrondo's paradox has been described in papers \cite{flitney-parrondo, abbott-paradox, parrondo-newgames, meyer-lga}. In this paper we present a new implementation scheme of a capital-dependent Parrondo's paradoxical games on a ~relatively small number of qubits.

\section{Parrondo's Paradox}
\subsection{Classical version}

Parrondo's paradox consists of a~sequence of games, where each game can be interpreted as a~toss of an asymmetrical coin. Every success means that the player gains one dollar, every loss means that the player loses one dollar. There are two games. Game $\A$ has probability of winning $1/2-\epsilon$. Game $\B$ depends on the amount of capital accumulated by player. If his capital is a~multiple of three, the player tosses coin $B_1$, which has probability of wining $1/10-\epsilon$, otherwise the player tosses coin $B_2$ which has probability of wining $3/4-\epsilon$. Originally $\epsilon=0.005$, but generally it can be any small real number.

Both games $\A$ and $\B$ are biased and have negative expected gain. But when a~player has the option to choose which game he wants to play at each step of the sequence, he can choose such a~combination of games which allows him to~obtain positive expected gain.

It is known that sequences $(\A\B\B\A\B)+$ or $(\A\A\B\B)+$ give relatively high expected gain. This fact is known as \textit{Parrondo's paradox}. 

\section{Proposed Quantum Implementation}
\subsection{Overview}
In \cite{meyer-lga, flitney-parrondo, FAJ04} the quantum versions of Parrondo games were proposed. 
The scheme introduced in \cite{flitney-parrondo} realizes a~history-dependent version of Parrondo's paradox. Its disadvantage is that it needs a~large number of qubits to store the history of the games. 
On the other hand' the scheme by Meyer and Blumer \cite{meyer-lga} uses Brownian motion of particle in one dimension and does not consume large amounts of quantum resources. The scheme presented by Flitney, Abott and Johnson in \cite{FAJ04} is based on multi-coin discrete quantum history-dependent random walk.

The implementation of the capital-dependent quantum Parrondo's paradox introduced in this paper uses only $O(\log_2(n))$ qubits, where $n$ is the number of games played. This allows to perform simulation even when a relatively large number of games are played; for instance, if a strategy consists of five elementary games then 400 iterations require only 15 qubits.

\subsection{Implementation}
\subsubsection{Gates and parameters}
To implement games $\A$ and $\B$, three arbitrary-chosen one-qubit quantum gates $A$, $B_1$ and $B_2$ are used. 
Each gate is described by four real parameters and our scheme as a~whole is described by set of parameters:
\begin{itemize}
\item $\{\delta_A, \alpha_A, \beta_A, \theta_A, \delta_{B_1}, \alpha_{B_1}, \beta_{B_1}, \theta_{B_1}, \delta_{B_2}, \alpha_{B_2}, \beta_{B_2}, \theta_{B_2}\}$: real numbers describing gates $A, B_1, B_2$;
\item $\mathbb{S}$: strategy -- any sequence of games $\A$, $\B$;
\item $n$: size of $\ket{\$}$ -- outcome register (see below);
\item $\offset$: initial capital offset.
\end{itemize}

Each gate is composed of elementary gates as presented in Eq.~(\ref{equ:gate-comp}):
\begin{equation}
G(\delta_G, \alpha_G, \theta_G, \beta_G)=R_z(\beta_G)R_y(\theta_G)R_z(\alpha_G)Ph(\delta_G),
\label{equ:gate-comp} 
\end{equation}
where $G\in\{A, B_1, B_2\}$ and\\
{\small
  $ 
  Ph(\xi)=
    \left(
      \begin{array}{cc}
      e^{i\xi} & 0\\
      0 & e^{i\xi}
      \end{array}
    \right),
  R_y(\xi)=
    \left(
        \begin{array}{cc}
        \cos(\frac{\xi}{2}) & -\sin(\frac{\xi}{2})\\
        \sin(\frac{\xi}{2}) & \cos(\frac{\xi}{2})
        \end{array}
    \right), 
  R_z(\xi)=
    \left(
        \begin{array}{cc}
        e^{-i(\frac{\xi}{2})} & 0\\
        0 & e^{i(\frac{\xi}{2})}
        \end{array}
    \right)
  $.
}

\subsubsection{Registers}
The quantum register used to perform this scheme consists of three subregisters: 
\begin{itemize}
  \item $\ket{c}$: one-qubit register representing the coin,
  \item $\ket{\$}$: $n$-qubit register storing player's capital,
  \item $\ket{o}$: three-qubit auxiliary register.
\end{itemize}
Register $\ket{c}$ holds the state of the \textit{quantum coin}. Gates $A$, $B_1$ and $B_2$ acting on this register represent \textit{quantum coin tosses}. One should note that the register $\ket{c}$ does not store information about history of the games. 

After every execution of gates $A$, $B_1$ and $B_2$, the state of the register $\ket{\$}$ is changed according to the result of the \textit{quantum coin toss}. This register is responsible for storing the history of the games, that is, player's capital. 

Register $\ket{o}$ is an ancillary register, which one need to check if the state of the $\ket{\$}$ register is a multiple of three. At the beginning of the scheme and after the application of the games' gates this register is always set to $\ket{000}$.

\subsubsection{Games}
Games $\A$ and $\B$ are implemented using the conditional incrementation--decrementation (CID) gate and gates $A$, $B_1$ and $B_2$ described above. In addition, game $\B$ uses gate $mod3$.

Gate $mod3$ sets $\ket{o_1}$ and $\ket{o_2}$ registers to state $\ket{1}$ iff the $\ket{\$}$ register contains a~number that is a multiple of three: 
\begin{equation}
	mod3\ket{a}\ket{0}=\ket{a}\ket{a\ (\mod\ 3)}.
\end{equation}

\begin{figure}[htbp]%
\centering
\subfigure[]
{
\includegraphics[scale=0.8]{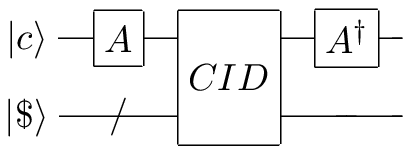}
\label{fig:game-a}
}
\subfigure[]
{
\includegraphics[scale=0.8]{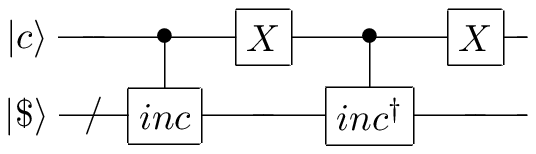}
\label{fig:cid}
}\\
\subfigure[]
{ 
\includegraphics[scale=0.8]{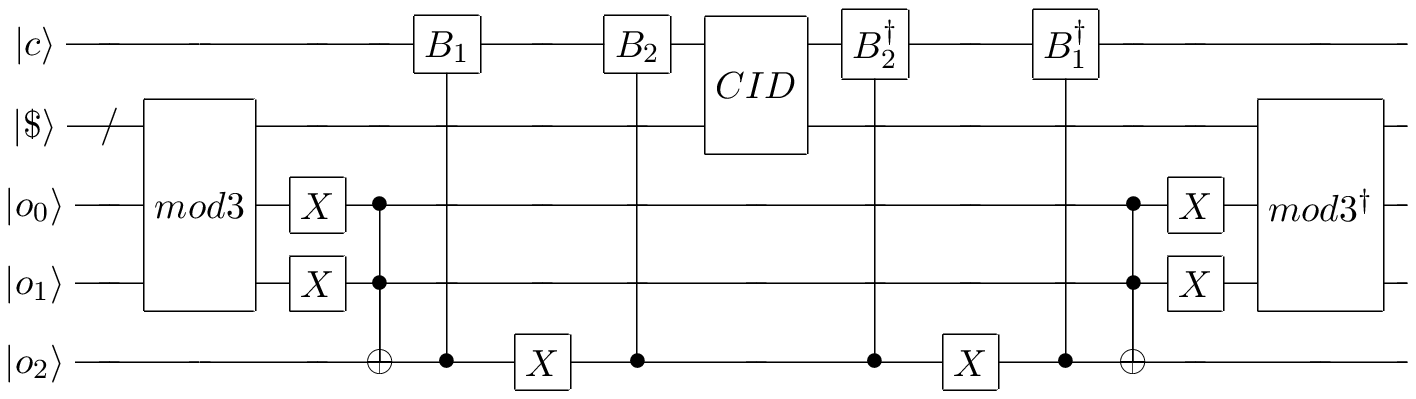}
\label{fig:game-b}
}
\caption{Gates used to implement Parrondo's paradox. (a) Conditional incrementation decrementation (CID) circuit. (b) Circuit for the game $\A$. (c) Circuit for the game $\B$.}
\label{fig:gates}
\end{figure}

The CID gate is responsible for increasing and decreasing the player's capital. The circuit for this gate is presented in~Fig.~\ref{fig:cid}. This gate increments register $\ket{\$}$ if $\ket{c}$ is in state $\ket{1}$ and decrements if it is in state $\ket{0}$.

Game $\A$ is directly implemented by gate $A$ as presented in Fig.~\ref{fig:game-a}.

Game $\B$, presented in Fig.~\ref{fig:game-b}, is more complicated. It uses gate $mod3$ to check if the player's capital is a multiple of three. If it is the case gate $B_2$ is applied to register $\ket{c}$, otherwise, $B_1$ is applied.

One can easily check that all gates used in this scheme are unitary because they are composed of elementary unitary operations.

\subsubsection{Sequence of games}
The game procedure is composed of the following steps:
\begin{enumerate}
\item Preparation of $\ket{c}$ in state $\frac{1}{\sqrt{2}}(\ket{0}+\ket{1})$.
\item Preparation of $\ket{\$}$ in state $\ket{(2^{(n-1)}+\offset)}$, where $\offset$ is a~small integer number,
\item Preparation of $\ket{o_1o_2o_3}$ in $\ket{000}$ state.
\item Application of gates $A$ and $B$ in some chosen order $\mathbb{S}$.
\end{enumerate}
After each application of gate $A$ or $B$ the number stored in register $\ket{\$}$ is either incremented or decremented. The initial state of register $\ket{\$}$ must be chosen in such way that integer overflow is avoided. The maximum number of elementary games cannot exceed the capacity of the register $\ket{\$}$.

\subsubsection{Outcome of games}
If our scheme is implemented on a~physical quantum device it should be finalized by measurement. This would give a~single outcome representing the final capital. Thus, to obtain expected gain, the experiment should be repeated several times. 

Simulation allows to observe the state vector of the quantum system. Using this property the expected gain is calculated as the average value of $\sigma_{z}$ in state $\ket{\$}\bra{\$}=\Tr_{\ket{c}\otimes\ket{o}}(\ket{c,\$,o}\bra{c,\$,o})$ obtained after tracing out the register with respect to coin and auxiliary subregisters:
\begin{equation}
	\langle\$\rangle=\Tr(\sigma_{z}^{\otimes n}\ket{\$}\bra{\$}).
\end{equation}

\section{Simulation}
Simulations of a quantum Parrondo's paradox presented in this article were performed using QCL \cite{omer-qcl}. The source code of the implementation can be found on the webpage listed in Ref. \cite{www:zksi}.


\subsection{Parameters}
To carry out the simulation, gates $A$, $B_1$ and $B_2$ were prepared with coefficients listed in Table~
\ref{tab:ex1}. Those coefficients were chosen arbitrarly. 

{ \renewcommand{\arraystretch}{1.2}
\begin{table}[!htbp]
\centering
\begin{tabular}{|c|c|c|c|}
\hline
\hbox{\vspace{1in} $\delta_A$ }& $\alpha_A$ & $\beta_A$ & $\theta_A$\\
  \hline 
0 & 1 & 0 & $2(\frac{\pi}{2}+0.01)$ \\
  \hline
  \hline 
$\delta_{B_1}$ & $\alpha_{B_1}$ & $\beta_{B_1}$ & $\theta_{B_1}$\\
  \hline 
0 & 1 & 0 & $2(\frac{\pi}{10}+0.01)$ \\
  \hline
  \hline 
$\delta_{B_2}$ & $\alpha_{B_2}$ & $\beta_{B_2}$ & $\theta_{B_2}$\\
  \hline
0 & 1 & 0 & $2(\frac{3\pi}{4}+0.01)$\\
  \hline

\end{tabular}
\caption{Coefficients of the experiment}
\label{tab:ex1}
\end{table}
}

\subsection{Results of simulation}
\begin{figure}[!htbp]
\centering
\subfigure[]
{
  \centering
    	\includegraphics[scale=0.850, angle=0]{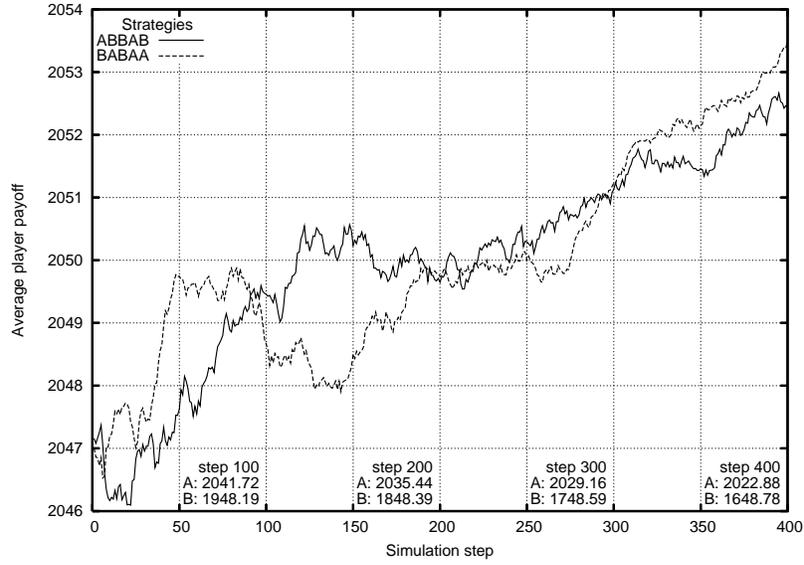}
  \label{fig:plot-o0}
}\\
    \subfigure[]
{
  \centering
    	\includegraphics[scale=0.850, angle=0]{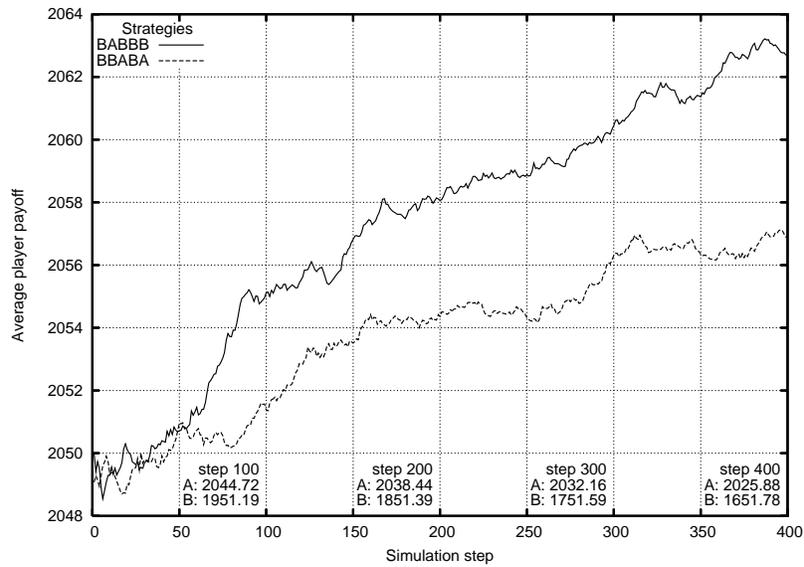}
  \label{fig:plot-o3}
}
\caption{Initial offset heavily influences the expected gain. For different offsets we have found different best strategies. The behavior of strategies $\A$ and $\B$ does not depend on initial $\offset$. (a) Comparison of two best-found winning strategies for $\offset=0$. Mean values for strategies $\A$ and $\B$ are also noted. (b) Comparison of two best-found winning strategies for $\offset=3$.} 
\label{fig:sim-res}
\end{figure}

In Fig.~\ref{fig:sim-res} the selection of results is presented. As one can see there are strategies that give positive expected values. For $\offset=0$, strategy $\A\B\B\A\B$ gives a~gain of $\sim5.43$ after 400 steps. For $\offset=3$, strategy $\B\A\B\B\B$ gives a~gain of $\sim13.69$ after 400 steps. 

\begin{figure}[!htbp]
	\includegraphics[scale=0.85, angle=0]{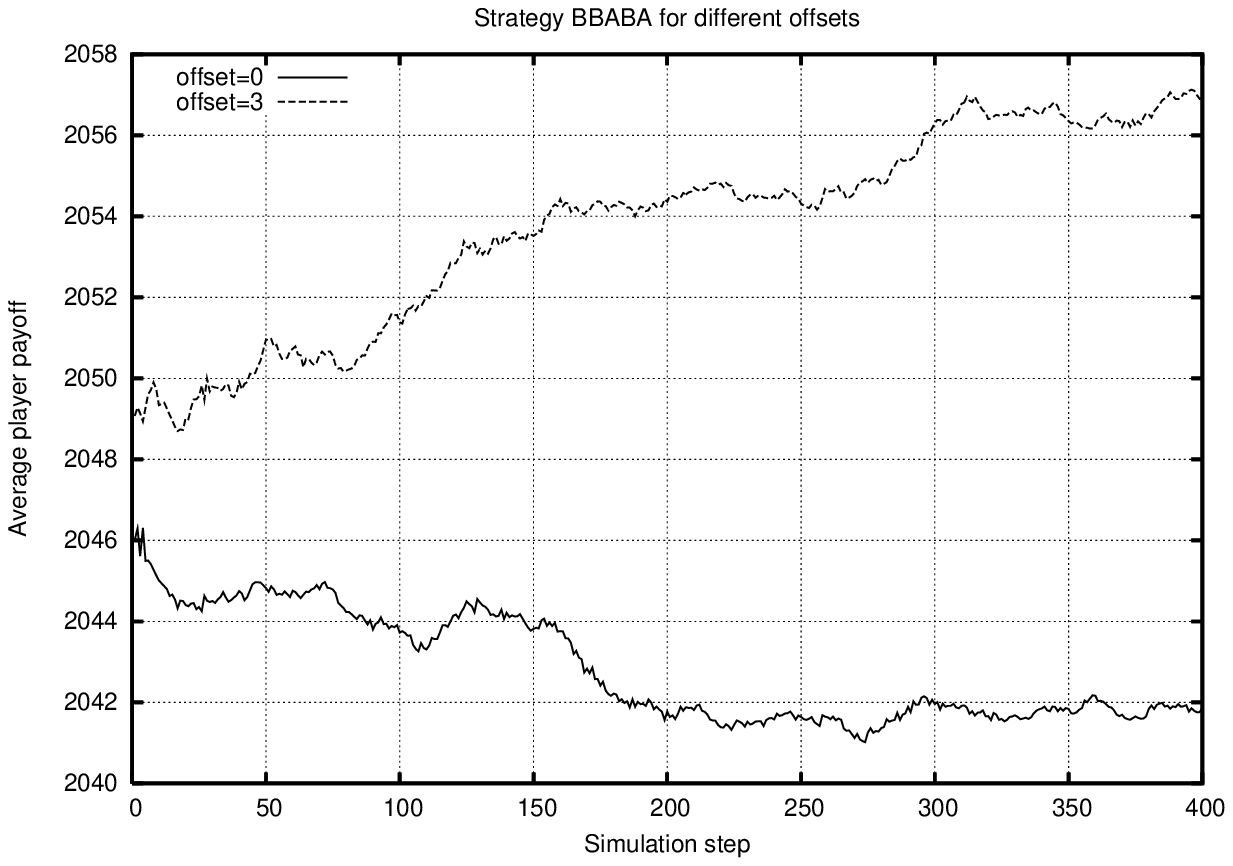}
	\caption{For different offsets the strategy can behave differently. In this case the strategy $\B\B\A\B\A$ is wining for $\offset=3$ and losing for $\offset=0$.} 
	\label{fig:sim-res-comp}
\end{figure}

Simulations have shown that finding the winning strategy for a~given initial set of parameters is not trivial because they are uncommon. We found that the initial value kept in register $\ket{\$}$ heavily influences the outcome, for example, see Fig.~\ref{fig:sim-res}. For different offsets different winning strategies can be found. One should note that behavior of the strategy can change if the initial offset is altered, for example see, Fig.~\ref{fig:sim-res-comp}.

\section{Conclusions}
We have shown that it is possible to create a capital-driven scheme for quantum Parrondo games using less than 20 qubits. The main advantage of this scheme is that the size of the register grows as $O(\log_2(n))$, where $n$ is the number of steps. We have found that the initial value of the register $\ket{\$}$ is important for selection of strategy. Simulations have shown that for an analyzed set of strategies composed of five elementary games, winning strategies are uncommon.

\section*{Acknowledgments}
This paper has been supported by the Polish Ministry of Scientific Research and Information Technology under the (solicited) grant No. PBZ-MIN-008/P03/2003.




\begin{thebibliography}{10}

\bibitem{PS03}
E.~W. Piotrowski and J.~S{\l}adkowski,
\newblock \textit{An invitation to quantum game theory},
\newblock {\em Int. J. Theor. Phys.}, {\bf 42} (2003) 1089.

\bibitem{PS04}
E.~W. Piotrowski and J.~S{\l}adkowski,
\newblock \textit{The next stage: quantum game theory},
\newblock in~{\em Mathematical Physics Research at the Cutting Edge}, Nova
  Science Publishers, Inc., (2004),
\newblock quant-ph/0308027.

\bibitem{flitabott-introduction}
A.~P. Flitney and D.~Abbott,
\newblock \textit{An introduction to quantum game theory},
\newblock {\em Fluct. Noise Lett.}, {\bf 2} (2002) R175.

\bibitem{Gro96a}
L.~Grover,
\newblock \textit{A fast quantum mechanical algorithm for database search},
\newblock in {\em Proc. 28th Annual ACM Symposium on the Theory of
  Computation}, ACM Press, New York (1996) pp. 212-219.

\bibitem{LeJo02}
C.~F. Lee and N.~Johnson,
\newblock \textit{Parrondo games and quantum algorithms},
\newblock quant-ph/0203043 (2002).

\bibitem{LeJo02-2}
C.~F. Lee and N.~Johnson,
\newblock \textit{Exploiting randomness in quantum information processing},
\newblock {\em Phys. Lett. A} {\bf 301} (2002) 343.

\bibitem{flitney-parrondo}
A.~P. Flitney, J.~Ng and D.~Abbott,
\newblock \textit{Quantum {P}arrondo's games},
\newblock {\em Physica A} {\bf 314} (2002) 35.

\bibitem{abbott-paradox}
G.~P. Harmer, D.~Abbott and P.~G. Taylor,
\newblock \textit{The paradox of {P}arrondo's games},
\newblock {\em Proc. R. Soc. Lond. ser. A} {\bf 456} (2000) 247.

\bibitem{parrondo-newgames}
J.~M.~R. Parrondo, G.~P. Harmer and D.~Abbott,
\newblock \textit{New paradoxical games based on {B}rownian ratchets},
\newblock {\em Phys. Rev. Lett.} {\bf 85} (2000) 5226..

\bibitem{meyer-lga}
D.~Meyer and H.~Blumer,
\newblock \textit{Parrondo games as lattice gas automata},
\newblock {\em J. Stat. Phys.} {\bf 107} (2002) 225.

\bibitem{FAJ04}
A.~P. Flitney, D.~Abbott and N.~F. Johnson,
\newblock \textit{Quantum random walks with history dependence},
\newblock {\em J. Phys. A} {\bf 37} (2004) 7581.

\bibitem{omer-qcl}
B.~Oemer,
\newblock \textit{Quantum programming in {QCL}},
\newblock Master's thesis, Technische Universit\"at Wien, 2000.

\bibitem{www:zksi}
http://www.iitis.gliwice.pl/zksi/.

\end{thebibliography}
\end{document}